\documentclass[sigconf,authorversion]{acmart}
\setkeys{acmart.cls}{balance=false}
\AtBeginDocument{%
  }

\copyrightyear{2025}
\acmYear{2025}
\setcopyright{cc}
\setcctype{by}
\acmConference[SC '25]{The International Conference for High Performance Computing, Networking, Storage and Analysis}{November 16--21, 2025}{St Louis, MO, USA}
\acmBooktitle{The International Conference for High Performance Computing, Networking, Storage and Analysis (SC '25), November 16--21, 2025, St Louis, MO, USA}
\acmDOI{10.1145/3712285.3759810}
\acmISBN{979-8-4007-1466-5/2025/11}

\newcommand{\tool}[0]{\mbox{\textsc{RAPTOR}}}

\newcommand{\flashx}{\mbox{Flash-X}}
\newcommand{\app}[1]{\textsc{#1}}
\newcommand{\fp}[0]{floating-point}
\newcommand{\flops}[0]{floating-point operations}
\newcommand{\flop}[0]{floating-point operation}
\newcommand{\capref}[1]{\S\ref{#1}}

\newcommand{\done}[1]{}

\usepackage[skip=.5\baselineskip]{subcaption}
\usepackage{fontawesome}
\usepackage{units}
\usepackage[capitalize]{cleveref}
\crefname{figure}{Figure}{Figure}

\usepackage{tabularx}                                                           
\newcolumntype{R}{>{\raggedleft\arraybackslash}X}                               
\usepackage{colortbl}
\usepackage{multirow}

\newcommand{\rC}[0]{\rowcolor[HTML]{EEEEEE}}                                    
\newcommand{\hC}[0]{\rowcolor[HTML]{333333}}

\usepackage{fancybox}
\cornersize*{2mm} 
\newcommand{\ourHypothesis}[2]{\\[3pt]\ovalbox{\begin{minipage}{\linewidth-2\fboxsep}\phantomsection\textbf{Hypothesis #1:} {#2}\end{minipage}}\\[0pt]}

\usepackage{calc}

\definecolor{m200}{RGB}{117, 235, 233}
\definecolor{m201}{RGB}{223, 170,  87}
\definecolor{m202}{RGB}{225, 112, 132}
\definecolor{m203}{RGB}{ 54, 114, 159}
\definecolor{m204}{RGB}{ 38,  87,  29}
\definecolor{m205}{RGB}{ 95,  20, 124}

\begin{document}

\title{RAPTOR: Practical Numerical Profiling of Scientific Applications}


\author{Faveo Hoerold}
\authornote{Faveo Hoerold and Ivan R. Ivanov contributed equally and are co-first authors.}
\email{fhoerold@student.ethz.ch}
\orcid{0009-0007-7472-5564}
\affiliation{
  \institution{ETH Zurich} 
  \city{Zurich}
  \country{Switzerland}                        
}
\additionalaffiliation{%
    \institution{RIKEN Center for Computational Science as Remote Trainee}
    \city{Kobe}
    \country{Japan}
}
        
\author{Ivan R. Ivanov}
\authornotemark[1]
\email{ivanov.i.e641@m.isct.ac.jp}             
\orcid{0000-0003-0356-3768}              
\affiliation{
  \institution{Institute of Science Tokyo} 
  \city{Tokyo}
  \country{Japan}                        
}
\additionalaffiliation{%
    \institution{RIKEN Center for Computational Science as Junior Research Associate}
    \city{Kobe}
    \country{Japan}
}
        
\author{Akash Dhruv}
\email{adhruv@anl.gov}             
\orcid{0000-0003-4997-321X}              
\affiliation{
  \institution{Argonne National Laboratory}
  \city{Lemont}
  \country{USA}
}

\author{William S. Moses}
\email{wsmoses@illinois.edu}          
\orcid{0000-0003-2627-0642}             
\affiliation{
  \institution{University of Illinois Urbana-Champaign}
  \city{Urbana}
  \country{USA}
}

\author{Anshu Dubey}
\email{adubey@anl.gov}          
\orcid{0000-0003-3299-7426}             
\affiliation{
  \institution{Argonne National Laboratory}
  \city{Lemont}
  \country{USA}
}

\author{Mohamed Wahib}
\email{mohamed.attia@riken.jp}          
\orcid{0000-0002-7165-2095}             
\affiliation{
  \institution{RIKEN Center for Computational Science}
  \city{Tokyo}
  \country{Japan}
}

\author{Jens Domke}
\email{jens.domke@riken.jp}             
\orcid{0000-0002-5343-414X}              
\affiliation{
  \institution{RIKEN Center for Computational Science} 
  \city{Kobe}
  \country{Japan}                        
}

\renewcommand{\shortauthors}{Hoerold et al.}

\begin{abstract}
The proliferation of low-precision units in modern high-perfor\-mance architectures increasingly burdens domain scientists. Historically, the choice in HPC was easy: can we get away with \unit[32]{bit} \fp{} operations and lower bandwidth requirements, or is FP64 necessary? Driven by Artificial Intelligence, vendors introduce novel low-precision units for vector and tensor operations, and FP64 capabilities stagnate or are reduced.
This forces scientists to re-evaluate their codes, but a trivial search-and-replace approach to go from FP64 to FP16 will not suffice.

We introduce \tool{}: a numerical profiling tool to guide scientists in their search for code regions where precision lowering is feasible. Using LLVM, we transparently replace high-precision computations using low-precision units, or emulate a user-defined precision. \tool{} is a novel, feature-rich approach---with focus on ease of use---to change, profile, and reason about numerical requirements and instabilities, which we demonstrate with four real-world multi-physics \flashx{} applications.
\end{abstract}



\begin{teaserfigure}
    \begin{center}
    \includegraphics{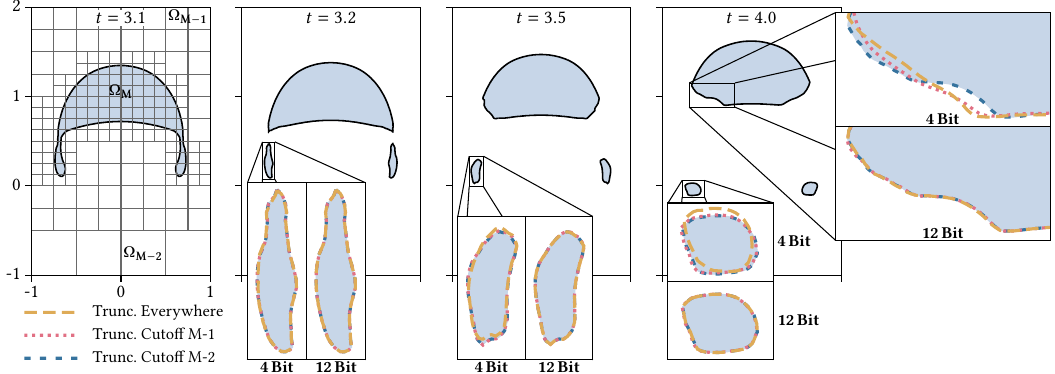}
    \caption{Qualitative visualization of bubble deformation and splitting over time ($t$) for the \app{Bubble} test case at Reynolds number $\text{Re} = 3500$, performed using incompressible multiphase fluid dynamics solvers in \flashx. The contours depict the zero level of the level-set function, $\phi = 0$, which defines the air-water interface ($\phi > 0$ in the air phase and $\phi < 0$ in the water phase). Adaptive mesh refinement (AMR) levels $\Omega_{M}$, $\Omega_{M-1}$, and $\Omega_{M-2}$ are shown and labeled in the left panel. AMR dynamically refines the mesh near the interface to accurately capture interface dynamics. We use \tool{} to truncate FP operations in the advection and diffusion modules of the Navier-Stokes solver following two different strategies. In the first strategy (\textit{Trunc.~Everywhere}), truncation is applied throughout the entire mesh. The second strategy selectively applies truncation in each mesh cell using a cut-off at the $M - l$ AMR level (where $l = 1, 2, \ldots$). Insets show zoomed-in views of the liquid-gas interface for simulations using \unit[4]{bit} and \unit[12]{bit} mantissas, highlighting the effect of reduced precision on the final simulation outcome. Further details in \capref{sec:fluidapp}.}
    \Description{Rising Bubble simulation at different times t=3.1, t=3.2, t=3.5, and t=4.0 with insets showing deviations of liquid-air interface for various truncation strategies in 4 and 12-bit simulations.}
    \label{fig:teaser}
    \end{center}
\end{teaserfigure}


\maketitle

\begin{table*}[tbp]
    \caption{Categorization and comparison of \tool{} to existing approaches and tools. These categories are used: (\textbf{A}) Algorithmic mixed precision, (\textbf{B}) Automatically changing precision, (\textbf{C}) System software-enabled precision changes, (\textbf{D}) Application-granularity, (\textbf{E}) Wrapper and emulator, (\textbf{F}) Precision format, and (\textbf{G}) Observe behavior w/o changing precision. Feature details are as follows: \textbf{Full app.~truncation} indicates support for unmodified software w/o infeasible runtime increase.\protect\footnotemark[2] \textbf{Dynamic truncation} means that truncation of variables can be made algorithmically dependent on the simulation state.\protect\footnotemark[3] \textbf{Flexible formats} stands for arbitrary precision support. \textbf{Scoped truncation} lets the user mark a function/region and the tool truncates the entire call stack below.\protect\footnotemark[4] \textbf{Granular truncation} means that individual operations can be in-/excluded from truncation.\protect\footnotemark[5] \textbf{Error tracking} traces errors through the application code.\protect\footnotemark[6] \textbf{Non-$\nabla$ Code} means that arbitrary (even non-differentiable) code is supported.} 
    \label{tbl:relwork}
    \centering
    {\footnotesize
        \begin{tabularx}{.84\linewidth}{llccccccccR}
        \toprule
        \multirow{2}{*}{\textbf{Approach}} & \multirow{2}{*}{\textbf{Category}} & \multicolumn{8}{c}{\textbf{Feature set}} & \multirow{2}{*}{\textbf{Ref.}} \\
        \cmidrule(lr){3-10}
        & & \begin{tabular}[c]{@{}l@{}}Full app.\\ truncation\end{tabular} & \begin{tabular}[c]{@{}l@{}}Dynamic\\ truncation\end{tabular} & \begin{tabular}[c]{@{}l@{}}Flexible\\ formats\end{tabular} & \begin{tabular}[c]{@{}l@{}}Scoped\\ truncation\end{tabular} & \begin{tabular}[c]{@{}l@{}}Granular\\ truncation\end{tabular} & \begin{tabular}[c]{@{}l@{}}Error\\ tracking\end{tabular} & 
        \begin{tabular}[c]{@{}l@{}}Non-$\nabla$\\ Code\end{tabular} &
        \begin{tabular}[c]{@{}l@{}}Supported\\languages\footnotemark[7]\end{tabular}\\
        \midrule \rC
        ADAPT     & B & \faThumbsODown & \faThumbsODown & \faThumbsUp & \faThumbsODown & \faThumbsUp & \faThumbsUp & \faThumbsODown & C, C++, Fortran & \cite{Menon2018} \\
        CADNA     & C & \faThumbsUp & \faThumbsODown & \faThumbsODown & N/A & \faThumbsUp & \faThumbsUp & \faThumbsUp & Ada, C, Fortran & \cite{Jezequel2021} \\ \rC
        FPSpy     & G & \faThumbsUp & \faThumbsODown & \faThumbsODown & N/A & N/A & \faThumbsODown & \faThumbsUp & Binary & \cite{Dinda2020} \\
        FPVM      & D & \faThumbsUp & \faThumbsODown & \faThumbsUp & \faThumbsODown & \faThumbsODown & \faThumbsUp & \faThumbsUp & Binary & \cite{Dinda2022} \\ \rC
        GPU-FPX   & G & \faThumbsUp & \faThumbsODown & N/A & N/A & \faThumbsODown & \faThumbsUp & \faThumbsUp & GPU Binary & \cite{Li2023} \\
        GPUMixer &  B & \faThumbsODown & \faThumbsODown & \faThumbsODown & \faThumbsODown & \faThumbsODown & \faThumbsUp & \faThumbsODown & CUDA & \cite{Laguna2019} \\ \rC
        Jost et al. & C, E, F & \faThumbsUp & \faThumbsODown & \faThumbsUp & N/A & \faThumbsUp & \faThumbsODown & \faThumbsUp & C & \cite{Jost2021} \\
        Gu et al. & E & \faThumbsODown & \faThumbsODown & \faThumbsODown & \faThumbsODown & \faThumbsUp & \faThumbsODown & \faThumbsUp & C, C++ & \cite{Gu2020} \\ \rC
        NEAT      & E & \faThumbsUp & \faThumbsUp & \faThumbsUp & \faThumbsODown & \faThumbsUp & \faThumbsODown& \faThumbsUp & Binary & \cite{Barati2021} \\
        Precimonious & A & \faThumbsODown & N/A & \faThumbsODown & \faThumbsUp & \faThumbsODown & \faThumbsODown & \faThumbsUp & C & \cite{Gonzalez2013} \\ \rC
        Puppeteer & D & \faThumbsUp & \faThumbsUp & N/A &  \faThumbsODown & \faThumbsUp & \faThumbsODown & \faThumbsODown & C, C++ & \cite{Parasyris2022} \\
        \midrule
        \textbf{\tool{}} & B, C, E  &\faThumbsUp & \faThumbsUp & \faThumbsUp & \faThumbsUp & \faThumbsUp & \faThumbsUp & \faThumbsUp & C, C++, Fortran & \\
        \bottomrule
        \end{tabularx}
    }
\end{table*}

\section{Introduction}\label{sec:intro}
Scientific computing has traditionally relied on double precision \fp{} arithmetic to ensure numerical accuracy and stability across a wide range of problems. However, the growing availability of hardware optimized for lower precision—such as single or even half precision—presents an opportunity to significantly reduce energy consumption and improve computational efficiency. The core dilemma lies in balancing performance gains with the need for reliable scientific results. Some numerical methods are more amenable than others to systematic reduction of precision and can rely on a robust mathematical foundation for maintaining confidence in the outcome. Of these, numerical linear algebra has witnessed the highest level of activity. Several other areas such as many ordinary- and partial-differential equation (ODE and PDE) solvers do not have such mathematically founded support for lowering precision. While many parts of a simulation may tolerate lower precision without compromising accuracy, identifying these regions is nontrivial and highly problem-dependent. Using full precision everywhere guarantees numerical robustness but can be wasteful, whereas an indiscriminate use of reduced precision risks introducing errors that propagate and distort the final outcome. \cref{fig:teaser} illustrates this for a \textit{rising bubble} simulation.

Here, the incompressible multiphase fluid-dynamics problem involves the deformation and splitting of an air bubble rising in a pool of water. \cref{fig:teaser} visualizes the evolution of the liquid-gas interface and highlights how different mantissa truncations and numerical precisions (\unit[4]{bit} vs.~\unit[12]{bit} mantissa) affect interface dynamics. Zoomed-in views emphasize these differences. A full analysis of the experiment is provided in Sections~\ref{sec:exp-design} to~\ref{sec:exp-results}. This example demonstrates that for such workloads the challenge is to develop intelligent strategies and tools that enable selective precision tuning without undermining the fidelity of scientific simulations. 

In this paper, we introduce a numerical profiling methodology and accompanying new tool, called \tool{}, that can numerically profile code execution and enable scientists to infer the impact of reduced precision in an actionable way. We apply \tool{} to a diverse set of 
application configurations in \flashx{}~\cite{Dubey2022}, a multiphysics multidomain application software, where mathematical guidance 
typically predicts that  double precision is needed almost everywhere. We use scientific intuition to selectively reduce precision in code components and observe the impact through the profiles generated with the help of \tool{}. These experiments effectively demonstrate a methodology where generated insights enable meaningful cost-benefit analysis for running simulation campaigns where mathematics does not provide any guidance. 

With focus on the scientific applications community, our paper makes the following contributions:
\begin{itemize}
    \item We introduce \tool{}\footnote{\url{https://github.com/RIKEN-RCCS/RAPTOR}} to easily profile and alter \fp{} precision in user-specified code regions for C, C++, and Fortran; including GPU and OpenMP support. [\cref{sec:fppt}]
    \item We devise a methodology and demonstrate how \tool{} can be used to experiment and reason about scientific application behavior to make informed choices about where to lower (or increase) the precision. [\cref{sec:exp-design}--\cref{sec:exp-results}]
    \item We discuss potential use cases of \tool{} in characterizing precision needs of HPC workloads for 
    hardware co-design in exploring the optimal distribution and the types of necessary \fp{} units. [\cref{sec:hardware-co-design}]
\end{itemize}

\section{Background, Motivation, and Related Work}\label{sec:background}
The exploration of mixed-precision arithmetic in scientific computing has gathered attention in recent years due to the potential to enhance computational performance and efficiency without compromising the accuracy of the results. By strategically combining different levels of precision within numerical algorithms, researchers aim to leverage the speed and reduced resource consumption of lower-precision computations while maintaining the numerical stability provided by higher-precision calculations.

Abdelfattah et al.~\cite{Abdelfattah2021} surveyed mixed-precision numerical linear algebra methods, discussing their theoretical foundations and practical applications. The study highlights how low-/mixed-precision techniques, e.g., iterative refinement, can achieve the accuracy of double-precision. Similarly, Higham et al.~\cite{Higham2022} provide an extensive review of mixed-precision algorithms in numerical linear algebra, covering a range of problems from factorization to solving linear systems. Bhola et al.~\cite{Bhola2024} conducted a comprehensive study on rounding error analysis for mixed-precision arithmetics. Their work provides both deterministic and probabilistic approaches to quantify accumulated rounding errors in various operations.  

While mixed-precision arithmetic has seen growing adoption in areas like linear algebra, machine learning, and some PDE solvers, several areas of numerical methods remain relatively underexplored because they involve complex numerical stability and error propagation behavior, making precision tuning less straightforward. Examples include methods such as multigrid (MG) and adaptive mesh refinement (AMR), nonlinear solvers and root-finding algorithms, general-purpose nonlinear solvers, spectral methods used in high-accuracy solutions of differential equations. These methods typically demand high precision throughout and have not been a focus of mixed-precision research due to the assumption that double precision is required for accuracy preservation.

It is, however, possible to leverage scientific and physical intuition to predict if---and where---it might be feasible to lower the precision. Since there is no robust mathematical framework to validate such assumptions, there is a need for tools that enable interactive and responsive exploration of truncation strategies and provide detailed and immediate insights into the impact of precision changes. 
%
We have conducted a review of the existing tools, and define six categories of what and how the approach/tool tackles the numerical analysis and a set of desired features.
Table~\ref{tbl:relwork} shows these categorizations and feature comparisons, but also highlights that none of the existing tools are feature-complete or easy to apply to an existing, potentially large code base and scientific workloads. 

\footnotetext[2]{E.g., \flashx{} can not realistically be analyzed with ADAPT without applying kernel extraction or similar techniques to isolate individual regions of interest in the workload.}
\footnotetext[3]{E.g., NEAT has functionality to change behavior based on the calling context.}
\footnotetext[4]{E.g., tools making use of LLVM IR have this functionality.}
\footnotetext[5]{E.g., the approach by Gu et al.~considered individual flops for tuning.}
\footnotetext[6]{E.g., GPUMixer demonstrated this by using shadow variables.}
\footnotetext[7]{\emph{Binary} for tools which are able to instrument executables.}
\setcounter{footnote}{7}

Only NEAT~\cite{Barati2021} offers a broad set of comparable features. However, it lacks the scoped precision changes which we rely on to analyze subsections of code or individual physics kernels in a multi-physics application.
ADAPT~\cite{Menon2018} can be used to collect \fp{} precision sensitivity profiles, but only works for differentiable applications.
This downside is missing from CADNA~\cite{Jezequel2021}, FPSpy~\cite{Dinda2020}, and FPVM\cite{Dinda2022}, but none of them support dynamic truncations (i.e., changes in mantissa length depending on the application's state or other compile-time or runtime conditions). 
Both GPU-FPX~\cite{Li2023} and GPUMixer~\cite{Laguna2019} exhibit similar shortcomings to the other approaches, on top of being exclusively applicable to GPGPU codes. 
The remaining tools we are aware of, e.g., the ones developed by Jost et al.~\cite{Jost2021}, Gu et al.~\cite{Gu2020}, as well as Precimonious~\cite{Gonzalez2013} and Puppeteer~\cite{Parasyris2022}, are unable to track numerical errors throughout the code (similar to our mem-mode, cf.~\cref{sec:mem-mode-debugging}) or correlate them to individual source locations.
\tool{} 
is designed to overcome these limitations.


\section{\tool{}: a Floating-Point Profiling Tool}\label{sec:fppt}
%
Our goal for \tool{} is to numerically profile applications or regions of interest to determine whether the workload is amenable to execution
with lowered precision throughout, under certain conditions only, or not all.
We aim to make the iterative process of making changes to an application based on a hypothesis, profiling, and confirming the results as frictionless as possible.

\subsection{Conceptual Design}
\begin{figure}
    \begin{subfigure}[b]{0.4\linewidth}
        \includegraphics[width=\linewidth]{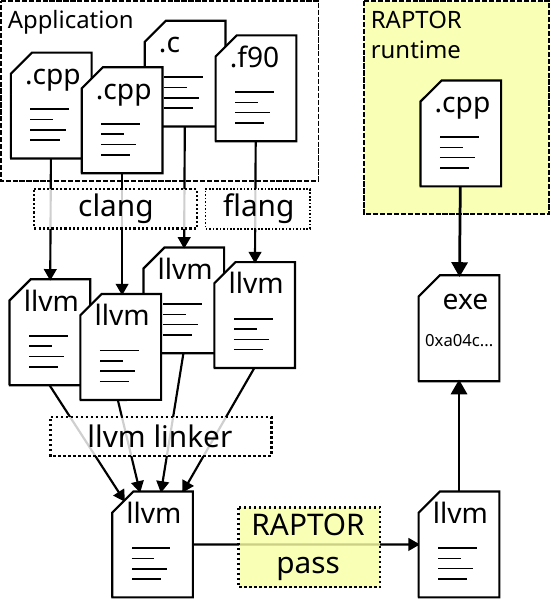}
        \caption{\tool{}'s components}
        \label{fig:raptor-overview}
    \end{subfigure}
    \hfill
    \begin{subfigure}[b]{0.485\linewidth}
        \footnotesize
        \begin{tabular}{rcc}
              
              Scope & op-mode & mem-mode \\ \hline
        Function & \cellcolor{green!20}\begin{tabular}[c]{@{}c@{}}Fully\\ Auto.\end{tabular} & \cellcolor{yellow!20}\begin{tabular}[c]{@{}c@{}}Semi\\ Auto.\end{tabular} \\ \hline
        File  & \cellcolor{green!20}\begin{tabular}[c]{@{}c@{}}Fully\\ Auto.\end{tabular} & \cellcolor{red!20}N/A                                                  \\ \hline
        Program & \cellcolor{green!20}\begin{tabular}[c]{@{}c@{}}Fully\\ Auto.\end{tabular} & \cellcolor{red!20}N/A                                                  \\ \hline
        
              && \\
              Features & op-mode & mem-mode \\ \hline
        Profile & \cellcolor{yellow!20} Local & \cellcolor{green!20}Extensive                                                  \\ \hline
        \begin{tabular}[c]{@{}c@{}}Precision \\ Increase \end{tabular} & \cellcolor{red!20} N/A & \cellcolor{green!20}Supported                                                  \\ \hline
        \end{tabular}
        \caption{Supported scopes and features for each mode}
        \label{tbl:raptor_modes}
    \end{subfigure}
    \caption{Overview of \tool{} and config.~matrix}%
    \Description{Left: Flowchart highlighting \tool{} components in a typical application build process. Right: Table comparing available scopes and features for op-mode vs. mem-mode truncation.}
    \label{fig:raptor-all}
\end{figure}

\paragraph{Configurations}
\tool{} can scope its transformation to three levels: function, file, and program; and it has two modes of operation: op-mode and mem-mode.
The supported configurations of scope and mode are shown in \cref{tbl:raptor_modes}.
The mem-mode requires more intervention from the user, and in turn, provides richer information about the computation (cf.~\cref{sec:raptor_modes} for further details).

\paragraph{Components}
\tool{} consists of two main components: a compiler instrumentation pass and a supporting runtime.
The compiler pass inserts calls to the runtime at appropriate places in the user's program.
The runtime executes \flops{} in the instructed precisions and collects data for analysis.
The placement of the components in the compilation pipeline is shown in \cref{fig:raptor-overview}.


\begin{figure*}
\includegraphics{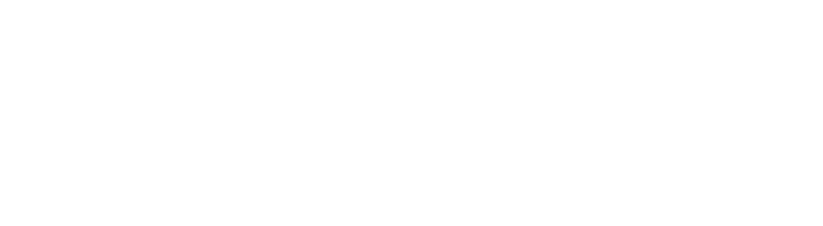}
\caption{Usage of \tool{} to go from original code to either op-mode (right) or mem-mode (left).}
\Description{Center: Plain C++ snippet with two functions foo and bar; foo calls bar. Right: C++ snippet demonstrating \tool{} op-mode wrapper for foo. Left: C++ snippet demonstrating tool{} mem-mode wrapper and function argument conversions for foo.}
\label{lst:raptor_op_usage}
\label{lst:raptor_pass_op_mode_before}
\label{lst:raptor_mem_usage}
\end{figure*}

\begin{figure}

\begin{subcaptiongroup}
\begin{minipage}{0.49\linewidth}%
\includegraphics{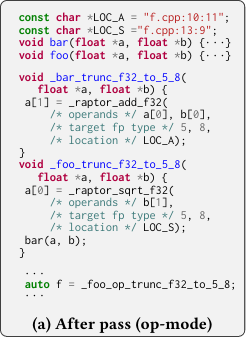}%
\phantomcaption%
\label{lst:raptor_pass_op_mode_after}%
\end{minipage}%
\hspace{1mm}%
\begin{minipage}{0.49\linewidth}%
\includegraphics{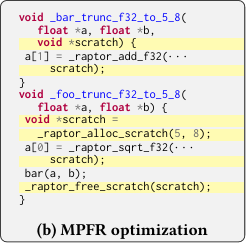}%
\phantomcaption%
\label{lst:raptor_mpfr_optimization}%
\end{minipage}%
\end{subcaptiongroup}

\caption{
\tool{}'s transformation pass replaces FP operations with runtime calls.
We also implement an optimization which allocates a temporary MPFR variable once to avoid re-allocation for each operation (cf. \cref{lst:raptor_runtime_impl_op}).
}
\Description{Left: C++ snippet showing \tool{} transformation replacing FP operations with runtime calls. Right: C++ snippet showing an optimization which avoids re-allocating memory for each MPFR operation.}
\label{lst:raptor_usage_and_transformation}
\label{lst:usage}
\end{figure}

\begin{figure}
\begin{subcaptiongroup}
\begin{minipage}{0.40\linewidth}%
\includegraphics{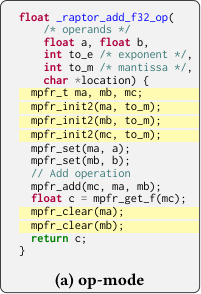}%
\phantomcaption%
\label{lst:raptor_runtime_impl_op}%
\end{minipage}%
\hspace{1mm}%
\begin{minipage}{0.585\linewidth}%
\includegraphics{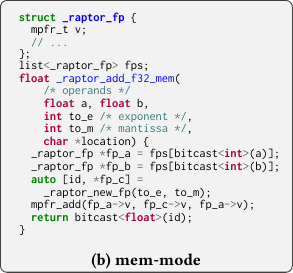}%
\phantomcaption%
\label{lst:raptor_runtime_impl_mem}%
\end{minipage}
\end{subcaptiongroup}
\caption{Implementation of the \tool{} runtime.}
\Description{Left: C++ snippet showing the \tool{} runtime implementation for the addition operation in op-mode. Right: C++ snippet showing the \tool{} runtime implementation for the addition operation in mem-mode.}
\label{lst:raptor_runtime_impl}
\end{figure}

\subsection{Usage}\label{sec:usage}
In general, there are flags\footnote{The flags differ by compiler/linker and are detailed in the paper's artifacts.} which must be added to load our plugin into the compiler and adjust its optimization pipeline.
The simplest way to use \tool{} is at the file or program scope.
This only requires an additional flag for the compilation command, e.g.: \verb|--raptor-truncate-all=64_to_5_14;32_to_3_8|.
The flag instructs \tool{} to perform a truncation on the currently compiled files for the instructed truncations (in this case \unit[64]{bit} width \fp{} operations to a type with \unit[5]{bit} exponent and \unit[14]{bit} mantissa, and analogously for \unit[32]{bit} width). If specified for all files, we achieve full-program truncation.
To use function scope truncation, the code needs to be instrumented as shown in \cref{lst:raptor_pass_op_mode_before}.
The user can request a truncated version of a function using the \verb|_raptor_trunc_func_[op,mem]| functions, depending on the desired mode. The parameters specify the function and \fp{} width to be truncated alongside the target exponent/mantissa lengths.\footnote{Currently, we require the exponent/mantissa to be compile-time constants.} The return value is a function with the same type as the input function.
%
Using mem-mode truncation, shown in \cref{lst:raptor_mem_usage}c, is more contrived and will be covered in detail in \cref{sec:mem-mode}.

\subsection{Instrumentation Pass}
The \tool{} pass is implemented as an instrumentation pass on the LLVM intermediate representation (LLVM IR)~\cite{Lattner2004}. LLVM is used by the clang (C/C++) and flang compilers among others. This allows us to handle all of Fortran, C, and C++ under a common representation suitable for compiler transformation and analysis.\footnote{We are providing examples in C++, however \tool{}'s concepts apply equivalently to C and Fortran as well (with syntactical differences).} We build on top of Enzyme's~\cite{Moses2020} infrastructure\footnote{Enzyme is an auto-differentiation framework for LLVM.} and make use of its instrumentation and plugin utilities.

\paragraph{Transformation}
\cref{lst:raptor_pass_op_mode_after} shows an example of the transformation when the user has requested a function scope truncation.
It is shown in C++ for illustrative purposes, but the actual code being transformed is in LLVM IR.
First, the pass finds all functions that the user requested be truncated and the truncation configuration (in this case, the user requested the function \texttt{foo} to be truncated). 
Then, the pass finds all transitively called functions (in this case, \texttt{foo} and \texttt{bar}), and for each of them, 
replaces all operations on \fp{} numbers 
with calls into the \tool{} runtime.
By utilizing features available in LLVM, we can recognize \fp{} arithmetic and functions in math libraries, such as the standard C and C++ math library.
All affected functions are cloned transformations in order to preserve the behavior of unrelated code that uses them.
When the requested scope is file or program-level, our pass applies the same transformation to the \flops{} of all functions, without the special handling required for function-scope truncation.
This is implemented as an LLVM pass plugin, 
which can be used in compilers such as clang or flang via their plugin interface in supported linkers (lld and gold), and as a standalone pass.

\paragraph{Preserving Compiler Optimizations}
The stage of the optimization pipeline in which the pass runs dictates how closely the truncated operations match the optimized version of the \flop{}. 
Compilers have various optimizations that affect \fp{} computation, e.g., vectorization or fusion.
However, they can only handle operations they are aware of.
For example, in \cref{lst:raptor_pass_op_mode_before}a, the code contains \fp{} arithmetic and a \texttt{sqrt} call to the libc math library, both of which are known to the optimizer, so it can treat them appropriately.
However, once the code has been transformed to call into our runtime (\cref{lst:raptor_pass_op_mode_after}), the optimizer no longer understands the semantics of the operation and no further \flop{} optimization is possible.

Ideally, our pass would run after all generic \fp{} optimizations are done to preserve as many compiler optimizations as possible. However, due to current limitations (see \cref{sec:limits})\done{enzyme? just not enough time to implement? can we get details?}, we are not able to handle vectorized instructions, which means we opt to run our pass before the vectorization passes.

\paragraph{Linking}
If a call to a function with no definition is encountered while attempting to truncate, we are not able to instrument any operations inside it.
For this reason, we configure the compiler(s) to use link-time optimizations (LTO), which merges all LLVM modules 
from different files into one file.
We run our pass after the merge, which allows us to analyze the entire call graph,\footnote{Calls to pre-compiled external libraries are ignored and \tool{} emits a warning.} see \cref{fig:raptor-overview}.


\subsection{Runtime}
\label{sec:raptor_runtime}
The \tool{} runtime executes \flops{} in the specified precisions and collects data for analysis.
This can be done using hardware, when the target precision is available on the CPU. Alternatively,
for arbitrary precisions, we use the GNU MPFR library~\cite{Fousse2007} for \fp{} emulation,
which provides feature rich and efficient implementations of arithmetic and mathematical functions.
\cref{lst:raptor_runtime_impl} sketches the MPFR-based wrappers in \tool{}.

The runtime also keeps track of how many \flops{} are executed and how much memory is accessed in truncated and non-truncated regions.
This information can be used to model an estimated speedup as shown in \cref{sec:hardware-co-design}.

\subsection{Operation Modes}
\label{sec:raptor_modes}

As mentioned earlier, \tool{} can operate in two modes: op-mode and mem-mode.
The main difference is how the runtime behaves.

\paragraph{Op-Mode}\label{sec:op-mode}

In this mode, the \fp{} values that cross the boundary between the \tool{} runtime and the user code (i.e. parameters to the runtime calls and return values) are represented using the pre-truncated \fp{} type.
For example, in the case in \cref{lst:raptor_pass_op_mode_after}, the arguments to the \verb|_raptor_add_f32| function and its return value will be valid \fp{} numbers of type \texttt{float}.
\cref{lst:raptor_runtime_impl_op} shows the runtime implementation. Each time we perform any \flop{} on these values, they first need to be transformed to the desired truncated precision (\verb|mpfr_set|), the operations performed in truncated precision, and then expanded again to the original \fp{} type (\verb|mpfr_get|).\footnote{This approach ensures correctness in adjunct code regions, e.g., calls to printf.}
%
%
Since the values are converted back to the original type after each operation, there is no way of tracking how values flow through the computation. 
So, this mode is 
useful for collecting statistics about errors in individual \underline{op}erations (hence, op-mode), and the number thereof.

As an optimization, to avoid expensive heap allocations for intermediate MPFR variables (\verb|mpfr_init2| in \cref{lst:raptor_runtime_impl_op}), we add a \emph{scratch pad} which is passed along as a parameter to functions in the truncated region.\footnote{This optimization is possible because \tool{} is implemented as part of a compiler, and hence we can alter call graphs and function signatures.}
This allows the runtime to allocate and free the temporary variables only once at the entry and exit of the truncated region (highlighted lines in \verb|_raptor_{alloc,free}_scratch| in \cref{lst:raptor_mpfr_optimization}), removing the need to execute the highlighted lines in \cref{lst:raptor_runtime_impl_op}.

\paragraph{Mem-Mode}\label{sec:mem-mode}
In this mode, values are not converted back to floats after each operation. Instead, the MPFR representation is \underline{mem}orized (hence, mem-mode) and is maintained between operations. Therefore, mem-mode allows precision increases, not only truncations. 

\cref{lst:raptor_runtime_impl_mem} shows a simplified version of the mem-mode runtime implementation. We use the bits in the \fp{} type to store an integer identifier, which is used to recover a struct (\verb|_raptor_fp|) containing the MPFR variable among other (customizable) bookkeeping data. 
%
Here, we add a double-precision shadow variable to the struct and update it with full-precision operations alongside the truncated MPFR variable. Hence, we can monitor the deviation between truncated variables and the (FP64) shadow variables for every operation, allowing us to set/monitor error thresholds (cf. \cref{sec:mem-mode-debugging}) and correlate them back to source code locations.
%

%
%
The additional capabilities in mem-mode come at an increased cost for user annotations, see \cref{lst:raptor_pass_op_mode_before}c compared to \cref{lst:raptor_pass_op_mode_before}a.
All variables (including arguments, return values, globals, values loaded or stored through memory) needed for the truncated region have to be converted to and from the new memory representation.\footnote{Our runtime offers converters: \_raptor\_\{pre,post\}\_convert (abbreviated in the figure), and we have implemented appropriate compiler warnings for potential regions.}

\begin{figure}[tbp]
    \centering
    \begin{subfigure}[b]{0.48\linewidth}
    \centering
    \includegraphics[width=\linewidth]{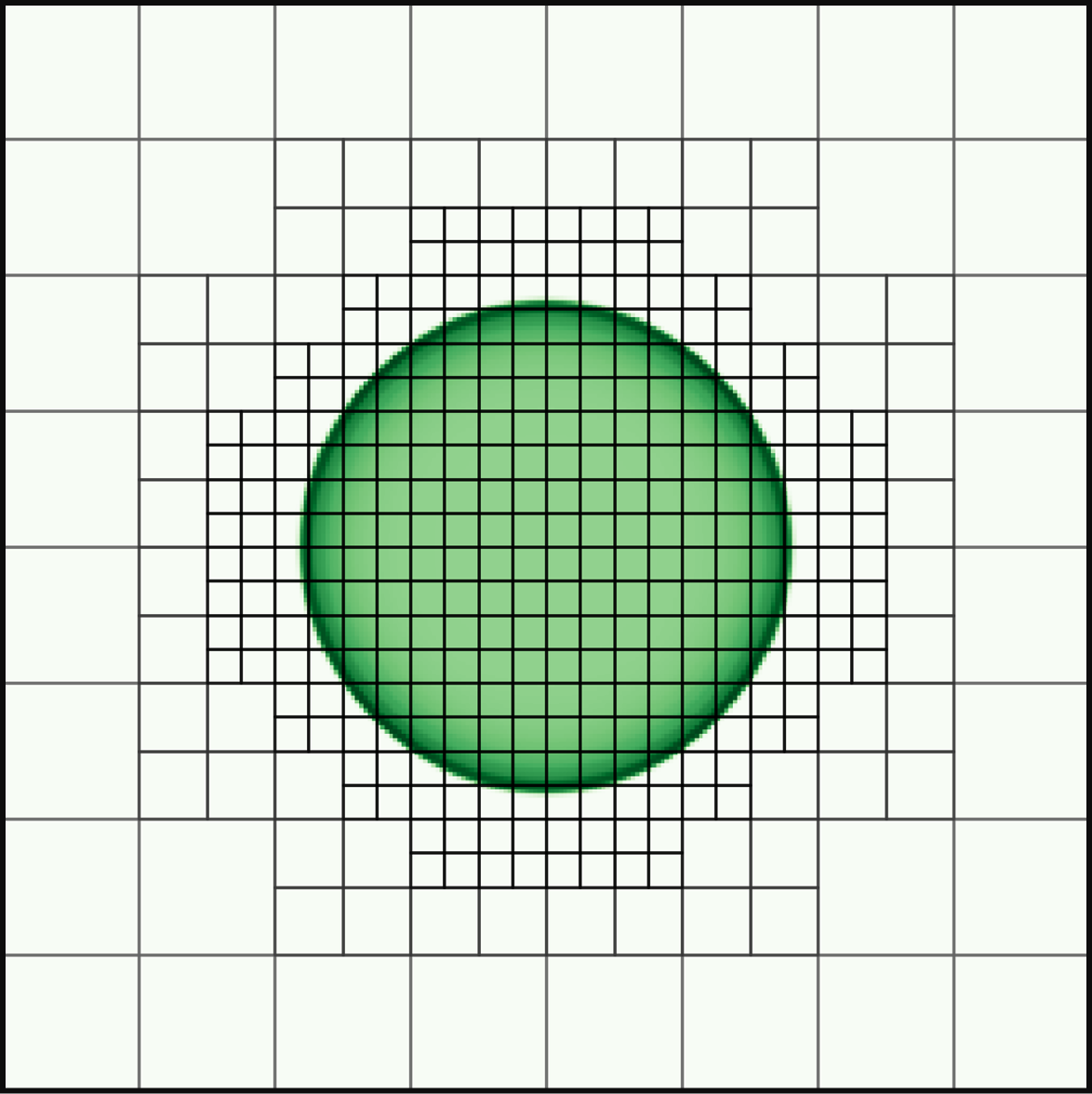}
    \caption{The radial shock in the Sedov blast simulation}
    \label{fig:sedov_shock}
    \end{subfigure}
    \hfill
    \begin{subfigure}[b]{0.48\linewidth}
    \centering
    \includegraphics[width=\linewidth]{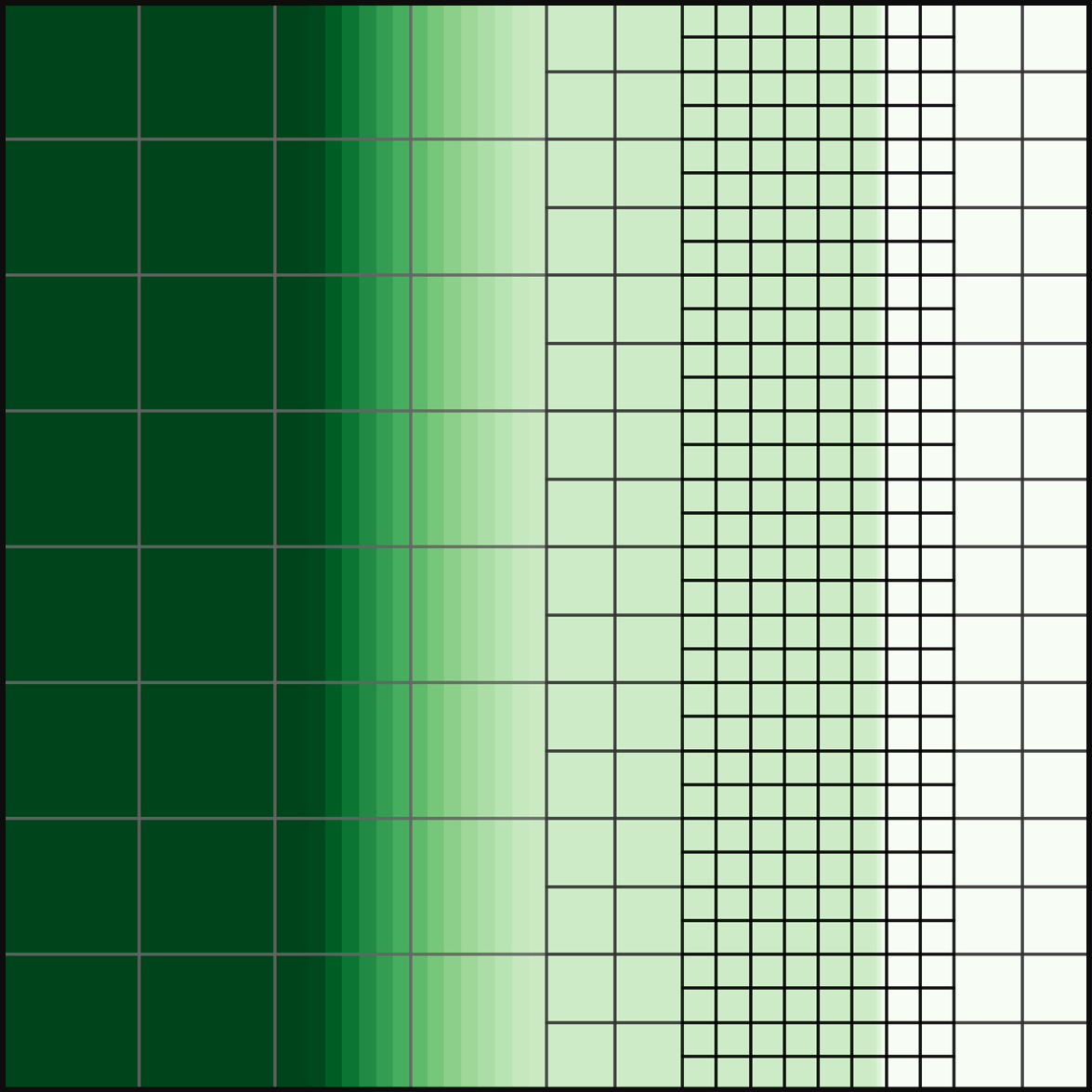}
    \caption{Shock along vertical plane in the Sod shock-tube simulation}
    \label{fig:sod_shock}
    \end{subfigure}
    \caption{Compressible hydrodynamics workloads with varying AMR block sizes. Each block contains a fixed number of mesh cells. The background color qualitatively represents pressure variations, with darker regions indicating higher pressure and lighter regions indicating lower pressure.}
    \Description{Visualizations of shock simulations with local mesh refinement near the shock front. Left: radial shock. Right: planar shock.}
    \label{fig:comphydro}
\end{figure}

\subsection{Compatibility}

\tool{} recognizes OpenMP directives and correctly truncates operations within nested OpenMP parallel constructs within the same device.

In addition to truncating applications for CPU, \tool{} can truncate code for GPUs (AMD and NVidia).
We have successfully tested OpenMP offloading and CUDA code.
Due to the lack of GPU support in arbitrary precision libraries (such as MPFR),
truncation is currently only possible to native types available in the GPU hardware, such as float or half precision \fp{}.

In general, \tool{}'s op-mode and MPI do not interfere with one another and truncation continues to work for any application with one or more MPI ranks. Most MPI operations only involve message passing and therefore require no special handling. However, \textsc{RAPTOR} does not implicitly truncate MPI reductions or any other MPI functions that perform \flops{}. If e.g. truncated MPI Allreduce is needed, a custom reduction operation can be implemented, which in turn can be truncated using \tool{}.
Mem-mode can only be used on shared-memory systems and without MPI reductions.\footnote{In op-mode, reductions do not get truncated automatically. In mem-mode, reductions will crash the application when they change the pointers to the shadow storage.}
Otherwise, additional MPI communication and handling would be required to transfer the shadow storage and adjust the pointers.

\section{Precision Considerations in Non-linear Multi-Physics Workloads}\label{sec:exp-design}


To evaluate \tool{} in the upcoming sections, we require a set of realistic application use cases. Our intent is to select a class of applications and numerical methods that do not have mathematical foundations that
could otherwise predict mixed-precision behavior.
We use \flashx~\cite{Dubey2022}, a scientific software that is used in multiple science domains. \flashx{} is
well suited for this study because its target applications are non-linear and coupled multi-physics problems which are sensitive to numerical perturbations. Additionally, \flashx{} uses Adaptive Mesh Refinement (AMR), a computational technique used to dynamically adjust the resolution of a simulation mesh based on the evolving features of the solution. Instead of using a uniform grid everywhere, AMR refines the mesh in regions where fine detail is needed (e.g., near steep gradients or discontinuities such as shocks and interfaces) and coarsens it where the solution is smooth. This allows simulations to achieve high accuracy while reducing computational cost by concentrating resources where they matter most.
Here, the use of
AMR permits us to exercise scientific intuition for adjusting the precision because AMR relies heavily on user-defined or computed thresholds for the refinement. 

\subsection{Experiment Design}
\flashx{} uses a version of AMR where the physical domain is divided into blocks that are spatially organized in an octree. 
The physical size of the blocks at a given level on the tree is identical; blocks one level above are twice the size along each dimension, while those one level below are half the size. Figure~\ref{fig:sedov_shock} illustrates this for a 2D domain. The number of cells is identical in every block.

Our first set of experiments (see \cref{sec:hydroapp} and \cref{sec:fluidapp}) is predicated on the intuition that the refinement criteria mentioned above ensure that all blocks that need the highest level of accuracy will be at the finest refinement level in the octree. The blocks at coarser levels in the tree will have smoother solutions implying that they may have a potential for adjusting the precision. In applications that simulate shock waves, the highest refinement will follow the progress of the shock, while in the applications which simulate multiphase flows, the highest resolution will follow the phase boundaries.
We can investigate
the impact of precision by varying the levels at which we commence the truncation, or by truncating specific physics solvers, and correlating the resulting error to our knowledge of
the physical regimes being explored in each simulation. These experiments use \tool{}'s op-mode with function and file-level scoping. 

With another set of experiments (see \cref{sec:mem-mode-debugging}) we will demonstrate the use of \tool{} where no prior assumptions are made based on scientific intuition. 
Here, we operate in a numerical debugging mode akin to performance debugging. We use mem-mode with function and file-level scoping for these experiments. In this mode, \tool{} flags the operations which deviate from a reference value by more than a predefined threshold. Working backwards from the flagged operations one can roll
back to full precision recursively until the desired accuracy is restored. 

\subsection{\flashx{} Workloads and Hypotheses}\label{sec:hypotheses}
\flashx{} can be configured for different applications with different underlying physics and solvers. Here, we select applications in two different physical regimes, i.e., with compressible and incompressible hydrodynamics solvers. 
In workloads with compressible hydrodynamics, shocks (sharp, nearly discontinuous change in the properties of fluid over a very small region of space) develop due to discontinuities in the initial conditions. We select two applications with different shock profiles. The first case, \app{Sedov} blast wave~\cite{sedov1993}, is initialized with a pressure spike in the
center of the domain where the shock moves out in the radial direction. The regions away from the shock are more or less quiescent (see Figure~\ref{fig:sedov_shock}). 
The second case is the \app{Sod} shock tube~\cite{Sod1978} where there is a jump in density along the vertical plane.
The shock wave moves in one direction while the rarefaction wave moves in the opposite direction, see Figure~\ref{fig:sod_shock}. For these workloads, we theorize:
\ourHypothesis{1}{\label{hyp1}In \app{Sedov}, some regions close to the shock have a higher resolution than the physical conditions strictly demand. Thus, we expect that reducing precision in all but the most refined blocks would not significantly impact the quality of the results. In \app{Sod}, the solution profile is less sharp and stretches across coarser blocks, and hence we expect reduced precision to have a larger impact on the results.}
%
%

Another study in the compressible regime is the cellular detonation~\cite{Timmes2000} (\app{Cellular}) with nuclear burning and an equation of state (EOS) suitable for stellar interiors.
The domain is initialized with pure carbon which is perturbed to ignite the nuclear fuel, producing an over-driven detonation that propagates along the $x$-axis. The EOS module uses a table of Helmholtz free energy with discrete values, and extrapolates them to match the conditions in the domain. For \app{Cellular}, we intend to explore the possibility of using lower precision in a solver other than \app{hydro} in a multiphysics scenario:
\ourHypothesis{2}{\label{hyp2}The EOS, used in the simulation, is table-based and is therefore the most likely candidate for reducing precision.}

The application in the incompressible regime is a rising bubble benchmark (\app{Bubble}). This solver employs a fractional-step projection method to evolve the velocity field and a sharp-interface ghost fluid method to model multiphase interfacial dynamics \cite{Dhruv2024,Dhruv2020}.
Here, a circular air bubble of diameter $d=1.0$ is initialized with its center at the origin, $(0, 0)$, within a two-dimensional rectangular domain. The air–water interface is tracked using a level-set function, $\phi$, where $\phi > 0$ denotes the air phase, $\phi < 0$ corresponds to the water phase, and $\phi = 0$ defines the air-water interface.

The \app{Bubble} simulation is governed by key dimensionless parameters for fluid and flow configuration. With $\rho'=1000$ and $\mu'=100$, we denote the density and viscosity ratios between water and air, respectively. These two parameters control the scaling of the fluid forces between the two phases. Furthermore, the Reynolds number ($\text{Re}$), Froude number ($\text{Fr}$), and Weber number ($\text{We}$) characterize the relative importance of inertial, gravitational, and surface tension forces. These numbers are set for water phase and scaled in air by the density and viscosity ratios, and we use $\text{Fr} = 1$ and $\text{We} = 125$. This configuration corresponds to a benchmark case reported in~\cite{Dhruv2024,Dhruv2023}.

We use the solution at $t=3$ computed at $\text{Re} = 35$ as the starting point for a series of simulations in which truncation is applied to both advection and diffusion operators of the Navier-Stokes solver. The advection terms are discretized using a fifth-order Weighted Essentially Non-Oscillatory (WENO) scheme, while a second-order central difference scheme is used for diffusion. These simulations are performed at $\text{Re} = 3500$. The choice of higher $\text{Re}$ for truncation tests was motivated by the need to accelerate bubble splitting and deformation in a shorter time period from $t=3$ to $t=4$. Our working hypothesis for this \app{Bubble} test is as follows:
\ourHypothesis{3}{\label{hyp3}The required precision will align with the AMR refinement strategy, specifically that reducing precision in lower-resolution blocks will not significantly degrade the overall quality of the simulation. We also expect that numerical precision will influence the evolution of the bubble interface, affecting deformation, splitting, and shape over time. Since the precision in velocity calculations directly impacts bubble dynamics, we expect noticeable differences as the precision varies.} 

\section{Experimental Setup}\label{sec:exp-setup}
%
Here, we briefly discuss versions and details of the sub-components of \tool{}. We also describe the execution environment and lay out our strategy for running the experiments.


%
\begin{figure*}[tbp]
  \centering
  \begin{subfigure}[b]{0.485\linewidth}
  \includegraphics{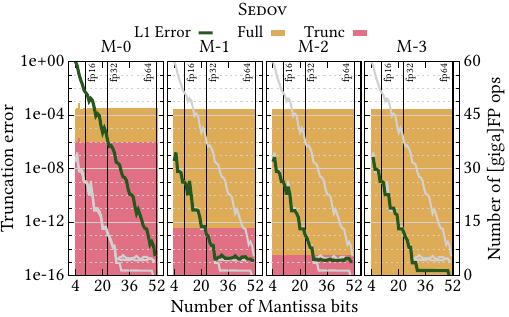}
  \caption{Between the 1st and 2nd panels, the error drops by seven orders of magnitude
    for mantissas smaller than \unit[18]{bits} and remains constant at $10^{-15}$ for
    mantissas larger than \unit[18]{bits}. Restricting truncation to coarser levels
    in the 3rd and 4th panels does not significantly change the error plot.}
  \label{fig:sedov}
  \end{subfigure}
  \hfill
  \begin{subfigure}[b]{0.485\linewidth}
  \includegraphics{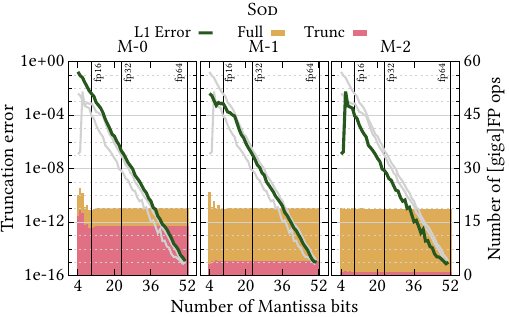}
  \caption{Between the 1st and 2nd panels, the error drops slightly for
    mantissas smaller than \unit[18]{bits} and matches the baseline elsewhere.
    In the 3rd panel, the error improves by half an order of magnitude
    for mantissas smaller than \unit[48]{bits}. The sharp error drop 
    for small mantissas is caused by the AMR algorithm refining all blocks.}
  \label{fig:sod}
  \end{subfigure}
  \caption{Truncating hydrodynamics in the \app{Sedov} blast wave simulation (left) and
    \app{Sod} shock tube simulation (right). The
    primary $y$-axis shows the $L1$ error norm of the fluid density compared to
    the full-precision simulation, as computed by \flashx{}' serial output comparison
    utility \app{sfocu}. Results are plotted with lines (not points) for clarity.
    Gray lines, showing the errors from the other panels, are included to emphasize small
    differences in error from one panel to the next. Black vertical lines mark half, float,
    and double precision mantissa sizes with 10, 23, and \unit[52]{bits}, respectively.
    In these plots, each panel represents a different refinement cutoff level beyond which
    truncation was disabled (i.e., $M-0$: truncate everything; $M-1$: disable truncation for
    most refined AMR level; $M-2$: disable truncation for two most refined levels; etc.).
    The bars in the background of each plot show the number of truncated operations (red)
    and full-precision operations (orange) stacked on top. As the refinement cutoff level
    is coarsened, the portion of truncated operations shrinks as truncation is restricted to
    coarser levels. \cref{sec:hardware-co-design} demonstrates how the operation counts
    can be used with a simple performance model to predict speedups.
    For small mantissas towards the left of each panel, the operation counts tend to deviate from
    the mean. The counts are perturbed when the AMR algorithm refines the mesh to compensate for
    inaccuracies caused by aggressive truncation. }
    \Description{Graphs showing simulation error induced by truncation, underlaid with stacked bar plots showing the proportion of truncated versus full-precision floating-point operations. Left: Sedov. Right: Sod.}
    \label{fig:compressible}
\end{figure*}


For LLVM, we use version 20, which offers a major improvement of the flang compiler over previous releases.
Our fork of Enzyme is based on the \emph{trunc-trace} branch.
We utilize Spack~\cite{Gamblin2015} to install MPFR version 4.2.1 and to install a \app{Bubble} dependency, called Hypre (v2.31.0).
We compile additional \flashx{} dependencies (HDF5 v1.14.6, OpenMPI v5.0.6, AmReX v24.08, MA28 v1.0.0) needed for the experiments.
To gain access to the compressible hydrodynamics applications, we fork \flashx{}' main branch, state of 2025/01/24.\footnote{\flashx{} carries an Apache 2.0 license but is maintained privately; Please email flash-x@lists.cels.anl.gov with your Github ID to get access (reference: \#ec2a750).} For the \app{Bubble} application, we fork \flashx{}' state of 2025/04/14.\footnote{For reference: \#cf6c018.}

We run all experiments in the default configuration for \flashx{}: CPU-only and MPI-only.
The \app{Sod}, \app{Sedov}, and \app{Cellular} applications execute sufficiently fast, and hence we use a single MPI process.
The \app{Bubble} workload is computationally more demanding, requiring us to execute it with 32 MPI ranks, while still confining it to a single compute node.
However, the parallelization across ranks does not affect the outcome of our experiments.\footnote{No MPI collectives are called within truncated sections; the domain is split over MPI processes and the physics routines that we truncate operate locally in each cell.}
Our cluster's compute nodes are equipped with dual-socket AMD EPYC 7773X processors and \unit[1024]{GiB} of DDR4 main memory. 

We employ the same methodology for each experiment: First, we evenly lower precision in a given module across the entire mesh to set the baseline of expected behavior. Then, we start with a very small mantissa and gradually increase its size getting an error estimate for each instance.  Once the baseline is established, we repeat the above process by lowering precision on blocks in the target physics solver at levels $M-1$, $M-2$, and $M-3$ respectively where $M$ is the maximum refinement level.

\section{\tool{} Case Studies: Experimental Results and Insights for the four AMR Workloads}\label{sec:exp-results}
%
%
Hereafter, we describe our results from analyzing the \flashx{} workloads with \tool{}. We explore three truncation modes: (1) Global truncation, where numerical values are truncated uniformly across the entire domain, (2) Selective truncation with AMR, where truncation is applied only on levels coarser than ${M-l}$, with $l$ controlling the cutoff depth in the AMR hierarchy, and (3) Selective truncation of a complete physics module. 

\subsection{Compressible hydrodynamics applications}\label{sec:hydroapp}
Figures \ref{fig:sedov} and \ref{fig:sod} show the results of the
\app{hydro} experiments.  As expected, the \app{Sedov} problem exhibits a very robust tolerance to lowered precision at $M-1$ as  
demonstrated in the second panel of \cref{fig:sedov},
which shows a much lower error
for any size of mantissa compared to the error shown in the first panel.
For mantissas with less than \unit[28]{bits}, excluding the finest AMR blocks
from truncation results in an error reduction of more than 7 orders-of-magnitude. For mantissas larger than \unit[28]{bits}, the error remains
constant and approaches the error measured for full truncation.
This error \emph{floor} appears as the simulation result drifts over time due to small
inaccuracies in truncated \fp{} operations.
Excluding the second level of AMR blocks from truncation does not
significantly improve the error of the simulation, as demonstrated
by the unchanging error in Panel~3. Only when we exclude the finest three
levels of AMR blocks from truncation do we see that the constant error floor drops
by another half order of magnitude as shown in Panel~4.

The colored bars in the background of the figure panels count the number of truncated and full-precision
\fp{} operations occurring for each simulation. In the fourth panel, the
workload is comprised almost entirely of full-precision operations, with truncated
operations making up less than $0.4\%$. 
In the third and second panel, the truncated operations make up progressively larger
portions of the workload as the Hydro module is truncated for more blocks in the domain.
Finally, in the first panel, the truncated operations make up more than 80\% of the
\fp{} workload. Additionally, for mantissas smaller than \unit[10]{bits}, the bars show
irregularities, with the overall flop counts fluctuating. This occurs because the
low-precision operations
are influencing the decisions of the AMR algorithm, which is responsible for dynamically
refining the mesh where needed. To be clear, it is not the algorithm itself which is
working with truncated precision. Rather, the algorithm notices imprecise blocks and decides
to refine them in order to maintain numerical accuracy and convergence in the simulation.

Figure \ref{fig:sod} shows the error plot for the \app{Sod} problem. Comparing the first two panels, we observe that excluding the finest AMR blocks from truncation results in an improvement 
in the overall error for mantissa sizes smaller than \unit[18]{bits}. This improvement results in an almost order-of-magnitude difference in error between the first
and second panel for simulations using \unit[4]{bits} of mantissa.
Panel 3 shows that excluding another level of AMR blocks from truncation results
in an error improvement across the board, with the difference starting at about
half an order-of-magnitude for large mantissas with \unit[48]{bits} and increasing to
almost an order-of-magnitude for small mantissas with \unit[10]{bits} or fewer.

Excluding refined blocks from truncation improves the error more for the
\app{Sedov} problem than it does for the \app{Sod} problem, validating Hypothesis \hyperref[hyp1]{1}. 
For \app{Sedov}, running just the finest blocks at full precision results in an
improvement of 7 orders-of-magnitude compared to truncating all blocks. For \app{Sod}, the same strategy decreases the error by an order-of-magnitude at most.
The figure for \app{Sod} has one less panel, because eventually no leaf blocks remain at the $M-3$ level in the simulation, so no truncation occurs and therefore there is no error.


In Figure \ref{fig:sod}, an anomalous behavior is visible for mantissas with \unit[4-6]{bits}, where
the errors are comparable to those obtained for mantissas with \unit[20]{bits}. The reason for that anomaly is written in the logfile of the \flashx{} run, i.e., it shows a higher number of leaf blocks (the blocks on which the solution evolves) which implies that many more blocks meet the refinement criterion with so few mantissa bits. Consequently, any advantage from lowering the precision further is more than offset by the increased number of refined blocks. 
It highlights the importance of a careful cost-benefit
analysis of any performance-related tweaks in the application
and showcases how \tool{} can assist with such an analysis.

With \app{Cellular} we turn our attention to the possibility of truncating an entire physics module. The EOS module is the obvious choice here because the ordinary differential equations (ODEs) in the Burn module are particularly stiff and sensitive to numerical perturbation. Hypothesis \hyperref[hyp2]{2} is based on the assumption that the computation can work with reduced precision because it extrapolates from a table look-up.
However, this hypothesis is proven to be incorrect because the Newton-Raphson algorithm, used to extrapolate from the table look-up values to actual conditions in the simulation, does not converge within the specified number of iterations when the mantissa is truncated to less than \unit[42]{bits}. In an attempt to get around this problem, we decrease the tolerance for convergence and increase the permitted number of iterations. Yet, we fail to get convergence for any meaningful workload, providing us with a counterexample for our hypothesis.
This experiment in particular highlights the usefulness of \tool{}
in numerical profiling because scientific intuition is not always reliable. Any direct modification of code to run such experiments with reduced precision would be time consuming and not as informative.
\subsection{Incompressible multiphase flow application}\label{sec:fluidapp}

For this experiment, we focus on the behavior of the bubble dynamics
under varying levels of precision. We omit the error curves because they do not provide any additional insights compared to the results already shown for
\app{Sedov} and \app{Sod} in \cref{fig:compressible}.
Figure \ref{fig:teaser} provides snapshots of the Rising Bubble benchmark at $\text{Re} = 3500$, showcasing the deformation and splitting of an air bubble rising through quiescent water. The interface is captured using a level-set formulation, with $\phi = 0$ denoting the liquid-gas boundary. The AMR hierarchy is centered around the interface to capture these fine-scale features with high fidelity. The key aim of this experiment is to test Hypothesis \hyperref[hyp3]{3}, by evaluating how different truncation strategies and numerical precision levels for the advection and diffusion operators affect the evolution of the bubble interface.

We compare results obtained with low (\unit[4]{bit}) and moderate (\unit[12]{bit}) precision. The zoomed-in insets in Figure \ref{fig:teaser} highlight the qualitative differences in interface topology under these settings. With aggressive truncation and low precision, the interface develops visible artifacts (see the $t=3.5$ and $t=4$ panel) particularly related to the shape of parent and satellite bubbles during and after break-up where the solution is highly sensitive to numerical inaccuracies. In contrast, with moderate precision and more selective truncation we are able to preserve the shape and physical accuracy of the interface evolution without requiring FP64 computations.

\subsection{Mem-mode debugging}\label{sec:mem-mode-debugging}
In this section, we demonstrate how the debugging output generated by \tool{} in mem-mode
can identify sections of code that may be unstable after truncation. A domain scientist
can use this information to understand which parts of the code contribute most to the overall inaccuracies
when truncated.

To recap \cref{sec:mem-mode}: mem-mode uses shadow variables that track a reference result for every operation as if the entire application had been run in FP64 up to that point.
This feature can be used to pinpoint exactly where calculations start to deviate from the reference (above a threshold).
%
\tool{} flags these occurrences of inexact operation results, groups them by (debug) location
in the code, and dumps the collected statistics when instructed by the user.
In practice, certain instances of operations get flagged more often than others
and flagged operations tend to be close together in the code. Thus, the output of mem-mode
builds a heatmap of code locations that do not react well to truncation.

We apply mem-mode debugging to the
\app{Sedov} problem with a newer, more modular hydrodynamics solver, Spark~\cite{Couch2021}.
The modularity of Spark makes it easier to ascribe a certain task to each part of the code.
For example, two important components of the solver are the {\em reconstruction} algorithm and
the {\em Riemann} solver. The reconstruction algorithm approximates the variation of the solution
within each cell with a profile instead of using a constant value.
The Riemann solver handles discontinuous solutions in shocks.
For this experiment, we keep the timestep of the solver constant to ensure that the dynamic
time-stepping algorithm does not compensate for inaccuracies resulting from truncation and thus skews our results.

%
\begin{table}[tbp]
    \caption{Numerically debugging \app{Sedov} with mem-mode}
    \label{tab:num-debug}
    \centering
    {\footnotesize
        \newcommand{\tblbetter}{\textcolor{ACMGreen}{$\blacktriangledown$}}
        \newcommand{\tblworse}{\textcolor{ACMRed}{$\blacktriangle$}}
        \begin{tabularx}{.95\columnwidth}{r|rrR}
            \toprule
            \multicolumn{1}{c}{} & \multicolumn{2}{c}{$L1$ error norm} & \\ 
            \multicolumn{1}{l}{Excluded modules} & \multicolumn{1}{c}{density} & \multicolumn{1}{c}{x-velocity} & \multicolumn{1}{c}{Truncated FP ops}\\ \hC
            \textcolor{white}{Baseline} & \textcolor{white}{8.113e-4} & \textcolor{white}{4.338e-3} & \textcolor{white}{90.6\%} \\ \rC
            Recon                  & \tblbetter{} 5.948e-4 & \tblbetter{} 3.258e-3 & 17.8\% \\
            Recon, Riemann         & \tblworse{}  9.300e-4 & \tblworse{}  6.929e-3 &  9.1\% \\ \rC
            Recon, Update          & \tblbetter{} 5.979e-4 & \tblbetter{} 3.180e-3 & 14.8\% \\
            \bottomrule
        \end{tabularx}
    }
\end{table}

We start by truncating the entirety of the hydrodynamics module and compute the resulting
errors for density and x-velocity using the \app{sfocu} utility, which is designed to verify the correctness of outputs from \flashx
~simulations against reference benchmarks (public known good solutions). These errors serve as a baseline for the rest of the experiment, as summarized in \cref{tab:num-debug}. The goal is to improve the error by selectively fencing off sections of the truncated module and running them at full double precision.

After we truncate the hydrodynamics, \tool{} flags a number of operations in the reconstruction algorithm. This can be interpreted in two ways. Either, (a) the reconstruction algorithm does not respond well to truncation and should therefore be run at full precision, or (b) another part of the code is responsible for introducing small \fp{} errors that only
get flagged when they are amplified in the reconstruction algorithm, for example due to multiplication or cancellation. If we exclude reconstruction from 
truncation (but still truncate
the remaining hydrodynamics code), then the errors for both density and x-velocity
decrease by a small amount. Next, a number of new operations get flagged in the Riemann solver and a single operation gets
flagged in the function responsible for updating the solution of the simulation. Adding the Riemann solver to the list of excluded functions drastically worsens errors, while
adding the update function leaves the errors essentially unchanged.

These results indicate that no specific part of the Spark solver is responsible for numerical sensitivity
in the \app{Sedov} workload and it is difficult to decide in advance whether a code section should be truncated or not.
Instead, an effective truncation strategy will dynamically turn on and off based on the local smoothness of the solution in the physical regime.
This small example reaffirms the need for a tool, such as \tool{}, that can help users to understand and reason about the parts of a code that may contribute the most to \fp{} errors.

\section{Discussion and Outlook}\label{sec:discussion}
We used \tool{} to test our hypotheses about how to truncate a number of
non-linear multi-physics workloads while keeping errors in check.
For \app{Sedov}, we found that restricting truncation to coarse blocks worked
well to keep errors low. Meanwhile for \app{Sod}, the same strategy resulted
in much higher errors. These two results match Hypothesis \hyperref[hyp1]{1}.
However, our Hypothesis \hyperref[hyp2]{2} was falsified when we tried to truncate
the EOS module in \app{Cellular} and found that it only converges when we run it with double precision. 
Finally, we evaluated the effect of precision on the evolution of the \app{Bubble} workload and found
that position and shape of the split bubbles depends on the strategy and level of truncation, as the scientific intuition
anticipated in our Hypothesis \hyperref[hyp3]{3}.

\subsection{Overhead}\label{sec:overhead}

\begin{table}[tbp]
    \caption{Slowdown of \tool{} in practice}
    \label{tab:perf}
    \centering
    {\footnotesize
        \begin{tabularx}{.95\columnwidth}{r|rrrrR}
            \toprule
            \multicolumn{1}{l}{} & \multicolumn{1}{c}{} & \multicolumn{2}{c}{Runtime (s)} & \multicolumn{2}{c}{Overhead (in $\times$)} \\ 
            \multicolumn{1}{l}{} & \multicolumn{1}{c}{Truncated FP ops} & \multicolumn{1}{c}{naive} &\multicolumn{1}{c}{opt.} & \multicolumn{1}{c}{naive} & \multicolumn{1}{c}{opt.} \\ \hC
            \multicolumn{6}{l}{\textcolor{white}{Sedov in op-mode (12 bit)}} \\
            M-0 & 86.3\% & 4781 & 1883 & 91.9 & 36.3 \\ \rC
            M-1 & 31.0\% & 1734 &  672 & 33.3 & 13.0 \\
            M-2 & 13.6\% & 794  &  323 & 15.3 & 6.2  \\ \rC
            M-3 & 0.37\% & 75   &   59 & 1.4  & 1.1  \\ \hC
            \multicolumn{6}{l}{\textcolor{white}{Sedov in op-mode with operation counting (12 bit)}} \\
            M-0 & 86.3\% & 5229 &   2336    & 24.2 & 45.0 \\ \rC
            M-1 & 31.0\% & 2088 &   1005    & 9.7 &  19.4 \\
            M-2 & 13.6\% & 1075 &    590    & 5.0 &  11.4 \\ \rC
            M-3 & 0.4\%  & 285  &    279    & 1.3 &   5.4 \\ \hC
            \multicolumn{6}{l}{\textcolor{white}{Sedov in mem-mode (12 bit)}} \\
            Truncate Hydro & 90.6\%   & 556 & N/A&148 &N/A\\
            Exclude Recon\protect\footnotemark  & 17.8\%   & 543 & N/A&147 &N/A\\
            \bottomrule
        \end{tabularx}
    }
\end{table}
\footnotetext{Exclusion in mem mode is handled dynamically in the runtime, so both entries have comparable overhead.}
\tool{}'s overhead for typical executions is shown in \cref{tab:perf}. The overhead
roughly correlates to the proportion of operations we truncate, and in the simplest
op-mode it can get up to 36$\times$ slower (with lower overheads for lower proportions).
Since our tool is not meant to be used in production runs, and is only meant for
experimentation and reasoning about precision, we believe that the overhead is within
acceptable ranges that make it useful. Additionally, in op-mode the overhead is almost entirely
rooted in MPFR's emulation of \flops{}. As mentioned in \cref{sec:raptor_runtime}, \tool{} can make use
of hardware types instead of MPFR emulation. In this case, we measure effectively zero
overhead, but of course only native types are available. In turn, mem-mode involves a large
amount of bookkeeping with a higher memory overhead and is thus more suited to analysis of
low-runtime sections.

\subsection{Use in Hardware Co-design}\label{sec:hardware-co-design}
We want to also highlight other adjunct usecases for \tool{}, e.g., as part of hardware co-design (w.r.t.~\fp{} capabilities).

\paragraph{Floating-Point Unit (FPU) Model}
We assume a simple model for a hypothetical processor with a fixed portion of chip
area dedicated to \fp{} processing.
Given that our investigated workloads are all double precision with some portion
truncated to a lower precision, we can assume that our CPU contains FPUs for double
precision and one lower precision. For simplicity, we assume that the areas dedicated
to each unit remain the same.


\begin{table}[tbp]
    \caption{Performance density of FPUs for various precisions (data from FPNew~\cite{Mach2020})}
    \label{tbl:fpu_density}
    \centering
    {\footnotesize
        \begin{tabularx}{.95\columnwidth}{lrrR}
            \toprule
            \multicolumn{1}{c}{FP Type} & \multicolumn{1}{c}{GFLOP/s} & \multicolumn{1}{c}{Area (kGE)} & \multicolumn{1}{c}{Perf. density (normalized)} \\ \midrule
            fp64 (11, 52)   & 3.17    & 53         & 1.00                            \\ \rC
            fp32 (8, 23)    & 6.33    & 40         & 2.65                            \\
            fp16 (5, 10)    & 12.67   & 29         & 7.30                            \\ \rC
            fp8  (5, 2)     & 25.33   & 23         & 18.41                           \\
            \bottomrule
        \end{tabularx}
    }
\end{table}

We estimate how much processing power in \fp{} operations per second (FLOP/s) a unit area of chip footprint provides for a given precision\footnote{For simplicity, we assume that floating point units with different precisions are completely independent. In reality, the units often share some components.}.
We get an estimate for the performance density of an FPU using data from an open-source RISCV FPU implementation (i.e., FPNew~\cite{Mach2020}), which is summarized in \cref{tbl:fpu_density}.
We extrapolate these values to get a performance density estimate for FPUs of any given precision.

Our theoretical CPU has a separate FPU for the two supported precisions (called $M_\mathrm{dbl}$ and $M_\mathrm{low}$), which, respectively, occupy an area of $A_\mathrm{dbl}$ and $A_\mathrm{low}$ and have a performance density of $P_\mathrm{dbl}$ and $P_\mathrm{low}$ operations per second per unit area (obtained above using extrapolation).
Therefore, the processing capability for a specific precision is bound by a peak of $A_i \times P_i$ operations per unit time. To estimate the area that each of the two FPUs will occupy, we assume a typical CPU configuration with double and single precision capabilities in a ratio of $1:2$ (e.g. Fugaku's A64FX~\cite{fugaku-specs}). Using the estimated performance densities $P_i$, and the ratio of the compute power, we get the ratio for the areas $A_\mathrm{dbl} : A_\mathrm{low} = 1.39$.

Next, we consider how to evaluate the time spent for \flops{} of a specific computation
which contains a number of \flops{} $N_i$ for each precision $i \in M$.
We assume that there is no parallelism across the FPUs and only a single type of FPU is in use at any given time, which matches the way we truncate all operations in specific regions in this work.

Consequently, the time spent for the computation in a precision~$i$ is $\frac{N_i}{A_i \times P_i}$ and the time spent for the entire computation is given by the following: 
$
    \sum_{i\in \mathrm{dbl}, \mathrm{low}}^{} \frac{N_i}{A_i \times P_i}
    \label{eq:flop_sum}
$.
We use this formula to obtain an estimated speedup for a computation in case it is compute-bound.


\paragraph{Memory Model}
As mentioned in \cref{sec:raptor_runtime}, we collect information about memory accesses that happen in truncated and non-truncated regions. We use these numbers to get a rough estimate of how much memory movement can be saved when \fp{} values are truncated to a lower precision. In case of memory-bound code, we assume that the runtime is a linear function of the required memory movement and obtain an estimated speedup.

\paragraph{Roofline Model}
To get a prediction for whether a computation is compute- or memory-bound, we build a roofline model of our processor, assuming a bandwidth of $1024$ GB/s (e.g., Fugaku's~\cite{fugaku-specs}). 

\begin{figure}[tbp]
 \centering
 \includegraphics{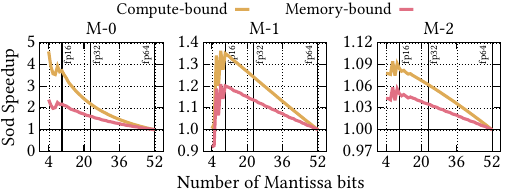}
 \caption{Estimated speedup of \app{Sod} (cf. \capref{sec:hypotheses} and \cref{fig:sod}) for different truncation strategies, according to our hardware model in compute-bound and memory-bound scenarios. Results are plotted with lines (not points) for clarity. Black vertical lines mark half, float, and double precision mantissa sizes with 10, 23, and \unit[52]{bits}, respectively. Irregularities around the small mantissas are due to AMR changing the ratios of truncated and non-truncated operations.}
 \Description{Graphs showing estimated compute-bound and memory-bound speedups of Sod for different truncation strategies.}
 \label{fig:sod_speedup}
\end{figure}

\paragraph{Results}
Using the above model and the data collected from our experiments described 
in \cref{sec:hydroapp},
we estimate speedups for different configurations of \app{Sod} and present the results in \cref{fig:sod_speedup}. This is a computation-heavy application with a
high operational intensity,
so our roofline model predicts a
compute-bound scenario. 
Predicted speedups for full truncation are
$3.7\times$
for
half precision and $2.2\times$
for
single precision ($2.2\times$, $1.6\times$ memory-bound).
The predicted
speedups for M-1 and M-2 are lower than for M-0 because a smaller portion
of the code is
truncated. Consequently,
speedups from low-precision operations impact the overall performance less.
Irregularities around the
small mantissas are caused by the adaptive behavior of AMR,
which changes the ratio of truncated and non-truncated operations.
This is reflected in the bar plots of
\cref{fig:sod}. For M-1, extra operations caused by AMR outweigh the speedup of
the low-precision operations, resulting in net slowdowns for
4 and \unit[5]{bit} mantissas.
With insight from domain scientists on acceptable levels of error obtained from
\cref{fig:sod}, this information can be used to obtain an optimal FPU configuration
for hardware co-design. Given that \tool{} simplifies the collection of such
information, collaborating with scientists for gathering data 
on the numerical behavior of software can become a powerful way to enable
supercomputing centers to make informed decisions 
about future procurements appropriate for their workloads.

\subsection{Current Limitations}\label{sec:limits}
For our proof-of-concept, we omitted a few quality-of-life features.
For example, not all elementary functions are implemented, but adding additional functions is trivial if MPFR already supports them. 
\tool{} does not truncate function calls to external, pre-compiled libraries\footnote{This can
be mitigated by making the function definitions available, either by providing source code or static libraries compiled with link-time optimization (LTO).} 
or calls to dynamic function pointers.
We have not tested thread-safety in general, 
original \fp{} operands have to be FP16, FP32, or FP64, and we do not support handwritten vector assembly. Limitations outside of supporting handwritten assembly can be overcome with small engineering effort.
As shown in \cref{lst:raptor_op_usage}, the user has to manually change a few lines of code to use scoped truncation. To improve the user experience, \tool{} can be enhanced for easier region annotation (e.g. \verb|#pragma trunc...|) and to support function filtering using a configuration file (similar to profilers).
Programming languages other than C/C++ and Fortran could be supported if they lower to LLVM IR.
Lastly, our truncation parameters are compile-time constants, but deciding the truncation level at runtime can be achieved by compiling multiple function pointers for different truncations and conditionally using them.

\section{Conclusion}\label{sec:concl}
This paper introduces a novel numerical profiling approach and accompanying tool,
called \tool{}, which overcomes the limitations of state-of-the-art tools in this
category. Using an LLVM compiler pass and MPFR, we are able to transparently
change the precision of \flops{} in selected
code regions. We create hypotheses for multiple, different workloads in the
\flashx{} framework for multi-physics simulations. Using our \tool{} tool, we
demonstrate how we are able to analyze and confirm/falsify these hypotheses based
on collected data for the non-linear solvers which are usually hard to reason about. 
Our hypotheses for \app{Sod}, \app{Sedov} and \app{Bubble} experiments prove to be correct. However, our intuition turns out to be incorrect for the \app{Cellular} experiment. 
Currently, we can apply our tool to C/C++ and Fortran codes for CPUs and GPUs, so it supports the majority of (large-scale) scientific workloads run on supercomputers. While we outline a few limitations, none of them are fundamental to our approach and we mention how each can be mitigated with additional engineering effort.
\tool{} is a step forward towards reasoning about mixed/low precision in scientific simulations, and we hope that it finds wide adoption in the scientific computing community.

\begin{acks}
This work was supported by the Scientific Discovery through Advanced
Computing (SciDAC) program via the  Office of Nuclear Physics and
Office of Advanced Scientific Computing Research in the Office of
Science at the U.S. Department of Energy.

This work was supported by \grantsponsor{jst_spring}{JST SPRING}, Japan Grant Number \grantnum{jst_spring}{JPMJSP2180} and the RIKEN Junior Research Associate Program.

This work was supported by the National Science Foundation under Grant
No. 2346519, and by the Alan Turing Institute, and the U.S. Department of Energy.

We gratefully acknowledge Hussein Harake and the Swiss National
Supercomputing Centre (CSCS) for providing access to computing resources
via the RACKlette student cluster, which was instrumental to this work.

\end{acks}

\bibliographystyle{ACM-Reference-Format}
\bibliography{references}

\end{document}